\documentclass[aps,prl,twocolumn,nofootinbib]{revtex4-2}
\usepackage{amssymb,amsmath}
\usepackage{epsfig,epsf}
\usepackage{bm} % puts greek and math symbols in boldface using \bm
\usepackage{color} % {\color{red} ... }
\usepackage{slashed}
\usepackage{hyperref}
\usepackage{tensor} %\indices{^a_b} produces the properly spaced tensor indices 
\newcommand{\beq}{\begin{equation}}
\newcommand{\beql}[1]{\begin{equation}\label{#1}}
\newcommand{\eeq}{\end{equation}}
\def\bal#1\gal{\begin{align}#1\end{align}}

\newcommand{\eq}[1]{(\ref{#1})}
\newcommand{\fig}[1]{Fig.~\ref{#1}}

\newcommand{\E}{{\rm e}}

\renewcommand{\b}[1]{{\bm #1}} 
\newcommand{\unit}[1]{\hat {{\bm #1}}} % unit vector

\newcommand{\aver}[1]{\left\langle #1 \right\rangle}
\begin{document}

\title{Synchrotron radiation by slowly rotating fermions}

\author{Matteo Buzzegoli}\affiliation{Department of Physics and Astronomy, Iowa State University, Ames, Iowa 50011, USA}

\author{Jonathan D. Kroth}\affiliation{Department of Physics and Astronomy, Iowa State University, Ames, Iowa 50011, USA}

\author{Kirill Tuchin}\affiliation{Department of Physics and Astronomy, Iowa State University, Ames, Iowa 50011, USA}

\author{Nandagopal Vijayakumar}\affiliation{Department of Physics and Astronomy, Iowa State University, Ames, Iowa 50011, USA}

\date{\today}

\pacs{}

\begin{abstract}
We study the  synchrotron radiation emitted by a  charged fermion, rotating as a part of a larger system, in a constant magnetic field $B$ parallel to the axis of rotation. The rotation is classical and independent of the magnetic field. The angular velocity of rotation $\Omega$ is assumed to be  much smaller than the inverse magnetic length $\sqrt{qB}$ which allows us to ignore the boundary effects at $r=1/\Omega$. We refer to such rotation as slow, even though in absolute value it may be an extremely rapid rotation. Using the exact solution of the Dirac equation we derived the intensity of electromagnetic radiation, its spectrum and chirality. We demonstrate by explicit numerical calculation that the effect of rotation on the radiation intensity increases with the particle energy. Depending on the relative orientation of the vectors $\b\Omega$ and $\b B$ and the sign of the electric charge, the rotation can either strongly enhance or strongly suppress the radiation. 

\end{abstract}

\maketitle

Synchrotron radiation is emitted by charged particles moving in magnetic fields. It has numerous applications in many areas of physics. In some systems the charged particles are a part of a larger rotating system that is subject to an external magnetic field. A computation of the combined effect of rotation and the magnetic field on the intensity of the electromagnetic radiation is the subject of this letter.  Our motivation derives from the recent discovery that the Quark-Gluon Plasma produced in relativistic heavy ion collisions possesses high vorticity \cite{STAR:2017ckg,Csernai:2013bqa,Deng:2016gyh,Jiang:2016woz,Xia:2018tes,Becattini:2015ska} and is subject to an intense magnetic field \cite{Kharzeev:2007jp,Skokov:2009qp,Voronyuk:2011jd,Bzdak:2011yy,Bloczynski:2012en,Deng:2012pc,Tuchin:2013apa,Zakharov:2014dia}. However, our results certainly apply to any rotating terrestrial or astrophysical system.

Consider a medium rotating in the laboratory frame with the constant angular velocity $\b\Omega=\Omega \unit z$. Let a fermion of electric charge $q$ and mass $M$ be embedded into the medium such that it is dragged by the medium to rotate with the same angular velocity. In particular, the medium exerts a radial force on the particle that balances the centrifugal force and prevents it from moving to infinity in the $(xy)$ plane. In the rotating frame, the only unbalanced force exerted on the particle is the Lorentz force due to  the constant magnetic field $\b B=B\unit z$. Thus, a classical particle performs a rotating motion about the $z$-axis with the synchrotron frequency. The particle trajectory in the laboratory frame can be obtained by rotating it through the angle $-\b\Omega t$. 

The quantum state of the fermion is described by the Dirac equation. In the frame rotating with the angular velocity $-\b\Omega$ we use the symmetric gauge $A^\mu=(0,-By/2,Bx/2,0)$, to cast it in the Schr\"odinger form  $i\partial_t\psi=H\psi$ with the Hamiltonian 
 \begin{align}\label{a4}
H=\gamma^0 \b\gamma\cdot(\b p-q\b A) + \gamma^0  + \Omega J_z\,,
 \end{align}
where  $\b p= -i\b\nabla$ and $J_z=-i\partial_\phi +\frac{i}{2} \gamma^x\gamma^y$ are the operators of momentum and longitudinal total angular momentum correspondingly and we use the units $\hbar=c=M=1$. The $\Omega$-independent part  $H_0$  of the Hamiltonian \eq{a4}  describes a fermion in the magnetic field and its spectrum $E_0$ and the corresponding eigenfunctions $\psi_0$ are well-known \cite{deOliveira:1962apw,Hehl:1990nf}. 

The leading order formula for the synchrotron radiation by a non-rotating\footnote{Throughout this letter, by `non-rotating' we mean that the system containing the fermion performs no rotation, i.e.\ $\Omega=0$.} fermion was obtained by Sokolov and Ternov \cite{Sokolov:1986nk}. In view of the axial symmetry, it is convenient to represent it in  cylindrical coordinates $\{t,\,\phi,\,r,\,z\}$. The functions $\psi_0$, corresponding to the non-rotating fermion, are eigenstates of $H_0$,  $p_z$, $J_z$ whose eigenvalues we denote as $E_0$, $p_z$ and $m$ correspondingly. While $p_z$ is continuous, the magnetic quantum number $m$ is a half-integer. Additionally, each eigenfunction $\psi_0$ is labeled by a  non-negative integer radial quantum number $a$. In a non-rotating system energy levels $E_0$ depend only on $p_z$ and the principal quantum number $n=m+\frac{1}{2}+a$. The energy eigenfunctions in the rotating frame can be obtained by replacing $E_0= E-m\Omega$: 
\begin{equation}
\label{eq:PsiSolutionGeneric}
\psi =e^{-i E t}\,
    \frac{e^{i p_z z}}{\sqrt{L}}  \frac{e^{i m \phi}}{\sqrt{2\pi}}\sqrt{|qB|}
     \left(\begin{array}{c}
         C_1 I_{n-1,a}(\rho) e^{-i\frac{\phi}{2}} \\
         i C_2 I_{n,a}(\rho) e^{i\frac{\phi}{2}} \\
         C_3 I_{n-1,a}(\rho) e^{-i\frac{\phi}{2}} \\
         i C_4 I_{n,a}(\rho) e^{i\frac{\phi}{2}}
    \end{array}\right)
\end{equation}
where $C_{1,3}=\frac{1}{2\sqrt{2}}B_+(A_+ \pm \zeta A_-)$,
 $C_{2,4}=\frac{1}{2\sqrt{2}}B_- (A_- \mp \zeta A_+)$
%\begin{equation}
%\label{eq:CoefficientsTransversePol}
%(C_1,C_2,C_3,C_4)=  \frac{1}{2\sqrt{2}}
%    \left( B_3 \left( A_3 + A_4 \right),
%     B_4 \left( A_4 - A_3 \right),
%     B_3 \left( A_3 - A_4 \right),
%     B_4 \left( A_4 + A_3 \right)\right)\,,
%\end{equation}
with $\zeta=\pm 1$ the fermion polarization and  
\begin{subequations}
\begin{align}
\label{eq:CoefficientsTransversePol2}
A_\pm = & \left(\frac{E-\Omega\, m \pm p_z}{E-\Omega\, m} \right)^{\frac{1}{2}},\\
B_\pm = & \left(1\pm\frac{\zeta}{\sqrt{(E-\Omega\, m)^2-p_z^2}} \right)^{\frac{1}{2}}.
\end{align}
\end{subequations}
We also defined $\rho = \frac{|q B|}{2} r^2$ and
\begin{equation}
\label{eq:def_I}
    I_{n,a}(\rho)=\sqrt{\frac{a!}{n!}}\E^{-\rho/2} \rho^{\tfrac{n-a}{2}} L_a^{n-a}(\rho)\,,
\end{equation}
where $L_n^{\alpha}(z)$ are the generalized Laguerre polynomials.
In \cite{Chen:2015hfc,Mameda:2015ria} the functions $\psi$ were obtained in a different form. The energy spectrum reads
\begin{align}\label{a8}
\left(E - \Omega\, m\right)^2 = 2 n |qB| + p_z^2 + 1\,.
\end{align}
Note that the energy levels explicitly depend on the magnetic quantum number $m$. 

The support of the wave functions $\psi_0$ is the entire Minkowski space. As a function of the radial distance $r$ from the rotation axis, they increase as a power law followed by exponential suppression. The typical size of the orbit can be gleaned from the expectation value $\aver{ r^2}= (2n+2a+1)/|qB|$. Unlike $\psi_0$, the wave functions $\psi$ belong to a rotating non-inertial frame. Causality demands these functions vanish at the radial distance $r=1/\Omega$ from the origin.\footnote{The importance of the causal boundary was discussed in Ref.~\cite{Duffy:2002ss,Ambrus:2015lfr,Ebihara:2016fwa,Chernodub:2017ref}} Nevertheless, if the magnetic field is strong and the rotation is relatively slow, the wave function $\psi$ is always exponentially small in the causality violating region and can be ignored there. More precisely, our derivation is valid as long as 
\begin{equation}\label{a39}
n,a\ll N_{\rm caus}\equiv\frac{|qB|}{2\Omega^2}\,.
\end{equation}

Consider now the photon wave function with given energy $\omega$, transverse momentum $k_\bot$, the longitudinal momentum $k_z$ and the magnetic quantum number $l$ in cylindrical coordinates: 
\begin{align}\label{PhotonWaveF}
\b A =\frac{1}{\sqrt{2\omega V}}
    \b\Phi e^{-i\omega t}\,.
\end{align}
We assume that photons are not interacting with the medium and, in particular, completely unaffected by the medium rotation.
It is convenient to choose $\b \Phi$ to be an eigenstate of the curl operator. The corresponding eigenfunctions are states with definite chirality:
\begin{equation}
\label{eq:TwistedPhotonWF}
\b\Phi=\frac{\omega}{k_\perp} \frac{1}{\sqrt{2}}
    \left(h\,\b T + \b P \right)e^{i\left(k_{z} z +l \phi\right)}\,,
\end{equation}
where $h=\pm 1$ labels right or left-handed photon states. 
The toroidal and poloidal functions appearing in \eq{eq:TwistedPhotonWF} read
\begin{align}
\b T&=\frac{i l}{k r} J_l\left(k_{\perp} r\right) \unit{r}-\frac{k_{\perp}}{k} J_{l}^{\prime}\left(k_{\perp} r\right) \unit{\phi}\label{eq:Toroidal}\\
\b P&=\frac{i k_z k_\perp}{k^2} J_l^{\prime}\left(k_{\perp} r\right) \unit{r}-\frac{l k_{z}}{k^2 r} J_{l}\left(k_{\perp} r\right) \unit{\phi} + \frac{k_\perp^2}{k^2} J_{l}\left(k_{\perp} r\right) \unit{z} .%e^{i\left(k_{z} z +l \phi\right)}\,.
\label{eq:Poloidal}
\end{align}
For a photon emitted at the polar angle $\theta$, $k_z= \omega \cos\theta$ and $k_\bot= \omega \sin\theta$. 

The photon emission amplitude by a fermion of charge $q$ transitioning between the two energy levels is given by the $\mathcal{S}$-matrix element
\begin{align}\label{c1}
\mathcal{S}=&(2\pi)\delta(E'+\omega-E)\frac{(-iq)}{\sqrt{2\omega V}}\\
&\times\int \bar\psi_{n',a',p_z',\zeta'}(\b x) \bm \Phi^*_{h,l,k_\bot,k_z}(\b x)\cdot \bm \gamma \psi_{n,a,p_z,\zeta}(\b x)\, d^3x\,,\nonumber
\end{align}
where primed quantities refer to the final energy level. Integrating and summing $|\mathcal{S}|^2$ over the phase space of the final particles, dividing it by the observation time and multiplying by the photon energy yields the differential radiation intensity for a photon with the circular polarization $h$:
\begin{widetext}
\begin{align}\label{eq:DiffIntensity}
   \frac{dW_{n,a,p_z,\zeta}^{h}}{d\omega} = &\frac{q^2}{4\pi}  \sum_{n',a',\zeta'}\delta_{m,m'+l}
   \int \omega^2\sin\theta  d\theta\, \delta(\omega-E+E') I_{a,a'}^2(x) \nonumber\\
&\times \Big|
\sin\theta\left[K_4 I_{n-1,n'-1}(x) - K_3 I_{n,n'}(x)\right]
 + K_1\left(h-\cos\theta\right)I_{n,n'-1}(x) - K_2\left(h+\cos\theta\right)I_{n-1,n'}(x)
\Big|^2\,,
\end{align}
\end{widetext}
where 
\begin{equation}
\label{eq:ABCD}
\begin{split}
    K_1 =& C_1' C_4 + C_3' C_2\,,\quad
    K_2 = C_4'C_1 + C_2' C_3,\\
    K_3 =& C_4' C_2 + C_2' C_4 \,,\quad
    K_4 = C_1'C_3 + C_3' C_1\, ,
\end{split}
\end{equation}
and we introduced a dimensionless variable $x=k_\bot^2/2|qB|$.
Conservation of the $z$-component of the angular momentum requires that $m=m'+l$. Energy conservation is expressed by the  
delta-function in \eq{eq:DiffIntensity} which can be written as 
\begin{align}\label{DeltaFunction}
\delta(\omega - E + E') 
=\frac{\delta(\omega - \omega_0)}{1+\frac{\omega\cos^2\theta}{E'-m'\Omega}}\,.
\end{align}
The characteristic frequency $\omega_0$ takes the simplest form in the frame where $p_z=0$, which we can always choose by virtue of the translation symmetry along the rotation axis:
\begin{align}\label{eq:PhotonEnergyRotation}
\omega_0 =& \frac{E-m'\Omega}{\sin^2\theta}
    \left\{ 1 - \left[ 1 - \frac{\mathcal{B} \sin^2\theta}{(E-m'\Omega)^2} \right]^{1/2} \right\}\,,
\end{align}
with 
\begin{align}\label{d11}
\mathcal{B}=& 2 (n-n') |qB| - \Omega^2 (m-m')^2\\
&+ 2 (E-m'\Omega) \Omega(m-m')\,. \nonumber
\end{align}

The radiation intensity for any initial and final fermion polarization states is obtained by summing over $\zeta'$ and averaging over $\zeta$: 
\begin{subequations}
\begin{align}
\overline{K_1^2} \equiv& \frac{1}{2}\sum_{\zeta,\zeta'} K_1^2
    =\overline{K_2^2} = \overline{K_3^2} = \overline{K_4^2} \nonumber\\
    =&\frac{(E-m\Omega)(E'-m'\Omega)-1}{4(E-m\Omega)(E'-m'\Omega)},\label{d25}\\
\overline{K_1 K_2} =& %\frac{1}{2}\sum_{\zeta,\zeta'} K_1 K_2
      \overline{K_3 K_4} 
    =\frac{\sqrt{2n|qB|}\sqrt{2n'|qB|}}{4(E-m\Omega)(E'-m'\Omega)},\label{d26}\\
\overline{K_1 K_4} =& %\frac{1}{2}\sum_{\zeta,\zeta'} K_1 K_4
     -\overline{K_2 K_3}
    =-\frac{\sqrt{2n|qB|}\omega\cos\theta}{4(E-m\Omega)(E'-m'\Omega)},\label{d27}\\
\overline{K_1 K_3} =& %\frac{1}{2}\sum_{\zeta,\zeta'} K_1 K_3
    \overline{K_2 K_4}= 0. \label{d28}
\end{align}
\end{subequations}

The total radiation intensity additionally requires  integration over $\omega$, which is trivial thanks to the delta function (\ref{DeltaFunction}):
\begin{widetext}
\begin{align}\label{I1}
   W_\text{tot}\equiv \frac{1}{2}\sum_\zeta W_{n,a,p_z=0,\zeta}^{h} = &\frac{q^2}{4\pi} \sum_{n',a'}\int_0^\pi d\theta\frac{\omega_0^2\sin\theta}{1+\frac{\omega_0\cos^2\theta}{E'-m'\Omega}}\frac{1}{2}\left(\Gamma^{(0)}_{n,a} + h \Gamma^{(1)}_{n,a}\right)
\end{align}
where 
\begin{align}\label{d30}
\Gamma^{(0)}_{n,a}= & I_{a,a'}^2(x)\Big\{
2 \overline{K_1^2}\left[ I_{n,n'-1}^2(x) + I_{n-1,n'}^2(x)\right]
    +\overline{K_1^2}\sin^2\theta\left[I_{n,n'}^2(x)
    +I_{n-1,n'-1}^2(x) - I_{n,n'-1}^2(x)-I_{n-1,n'}^2(x)\right]\nonumber\\
&-2\overline{K_1 K_2}\sin^2\theta \left[ I_{n,n'}(x)I_{n-1,n'-1}(x) + I_{n-1,n'}(x) I_{n,n'-1}(x) \right]\nonumber\\
&-2\overline{K_1 K_4} \sin\theta \cos\theta \left[I_{n-1,n'-1}(x)
   I_{n,n'-1}(x)+I_{n-1,n'}(x) I_{n,n'}(x)\right]\Big\}\\
    \label{d31}
\Gamma^{(1)}_{n,a}= & I_{a,a'}^2(x)\Big\{
    2 \overline{K_1^2} \cos\theta \left(I_{n-1,n'}^2(x)-I_{n,n'-1}^2(x)\right)\nonumber\\
&+2\overline{K_1 K_4} \sin\theta  \left(I_{n-1,n'-1}(x) I_{n,n'-1}(x)
    -I_{n-1,n'}(x) I_{n,n'}(x)\right)\Big\}.
\end{align}
\end{widetext}

Integration over $\theta$ as well as summation over $n'$ and  $a'$ can only be done numerically. The summations are restricted by energy and longitudinal momentum conservation,
and by the causality constraint \eq{a39} applied to the quantum numbers $n'$ and $a'$. A similar inequality was found in \cite{Chen:2015hfc}. Without the causality constraint \eq{a39}, which itself follows from $\aver{r^2}\Omega^2 < 1$, the sum over the phase space would be divergent. 
However, it can be shown that as long as $\Omega\ll\sqrt{|qB|}$, the intensity is independent of the cutoff $ N_{\rm caus}$. 
In the no-rotation limit $\Omega\to 0$, one can show that the sum $\sum_{n',a'}$ reduces to $\sum_{n'=0}^{n} \sum_{a'=0}^{\infty}$. Moreover, in this limit, the photon energy $\omega_0$ depends only on $n$ and $n'$, but not on $a$ and $a'$. This allows explicit summation over $a'$ in \eq{I1}, which can be performed  using the identity $\sum_{a'}I^2_{a,a'}(x)=1$ ~\cite{Sokolov:1986nk}  and yields the well-known result for the synchrotron radiation intensity by a non-rotating fermion. We numerically verified that our results are not sensitive to the cutoff $N_\text{caus}$.

%%%%
\begin{figure}[ht]
      \includegraphics[width=3.2in]{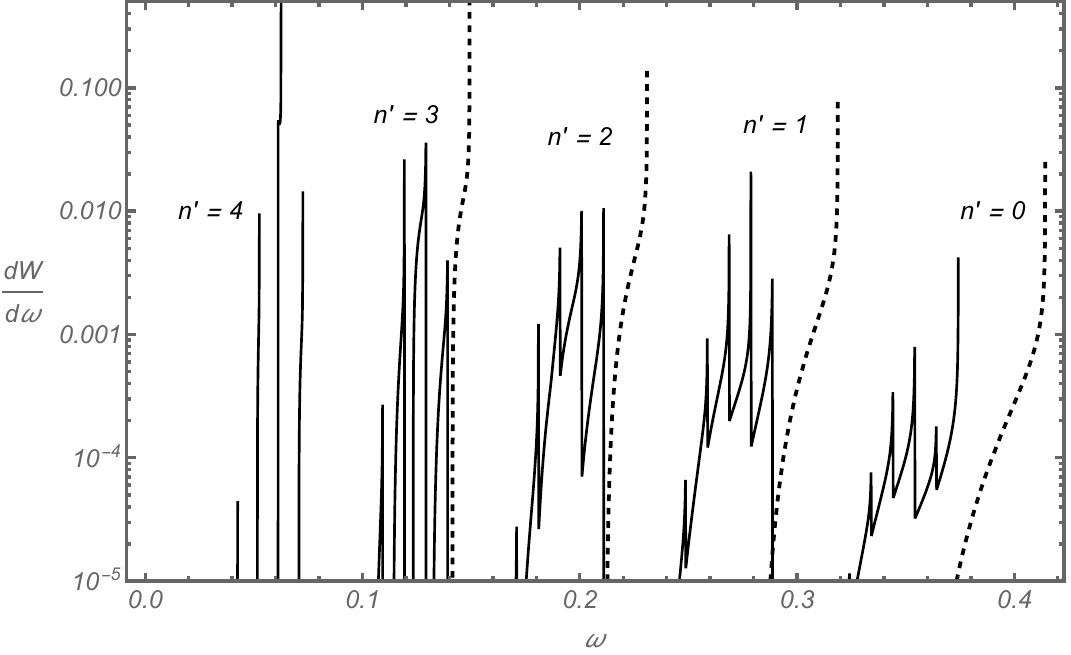}
  \caption{The spectrum of synchrotron radiation (\ref{eq:DiffIntensity}) at $qB=-0.1$ emitted by a fermion with initial quantum numbers $n=5$, $a=1$, $m=7/2$ and $p_z=0$, summed over $a'\in[0,16]$ and averaged over photon helicity $h$. Solid lines: $\Omega=-0.01$ (corresponding to $E=1.379$), dashed lines: $\Omega=0$ (corresponding to $E=1.414$). Our units: $\hbar=c=M=1$. }
\label{fig:spectrum}
\end{figure}
%%%%%

%%%%
\begin{figure}[ht]
      \includegraphics[height=8cm]{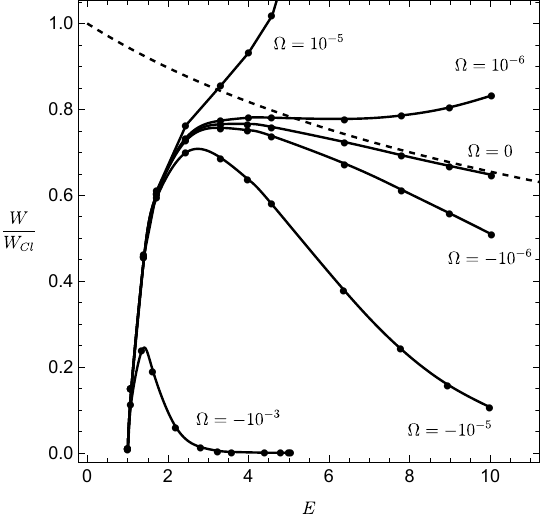}
  \caption{The total intensity of the synchrotron radiation in units of the classical intensity \eq{d35} as a function of the initial energy $E$ at $qB=-0.01$. Solid lines correspond to various angular velocities $\Omega$ and the dashed line is the quasiclassical approximation at $\Omega=0$. The dependence of the intensity on the initial value of $a$ is weak and not noticeable in the figure. Our units: $\hbar=c=M=1$. }
\label{fig:total1}
\end{figure}
%%%%%

%%%%
\begin{figure}[ht]
      \includegraphics[height=8cm]{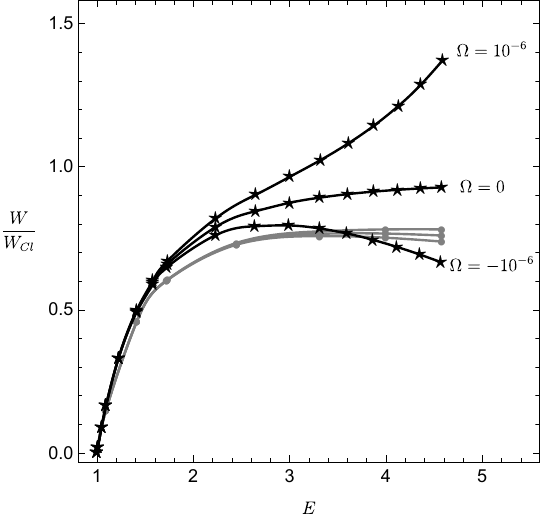}
  \caption{The same as in \fig{fig:total1} but with $\Omega = 0,\pm 10^{-6}$ and two values of $qB$. Grey lines with circles are for $qB = -10^{-2}$. Black lines with stars are for $qB =- 10^{-3}$. Note the effect of rotation is greater at lower energies for the smaller field. Our units: $\hbar = c = M = 1$.}
\label{fig:total2}
\end{figure}
%%%%%

A typical synchrotron radiation spectrum is shown in \fig{fig:spectrum}. For comparison we also plotted the spectrum emitted by the non-rotating fermion. It is  seen that while the spectrum of the non-rotating fermion depends only on the principal  quantum number $n'$, the spectrum of the rotating fermion is split in many lines having different $a'$, as expected from the energy shift caused by rotation. Moreover, the positions of the spectral lines are shifted toward smaller values of $\omega$ and their heights are diminished in comparison with the non-rotating spectrum. This indicates that the radiation intensity is suppressed when $\b B$ is anti-parallel to $\b \Omega$ and $q>0$.

The total intensity is conventionally represented with respect to the corresponding classical expression  
\begin{align}\label{d35}
W_\mathrm{cl}=\frac{q^2}{4\pi}\frac{2 (qB)^2E^2}{3}\,.
\end{align}
The result is shown in \fig{fig:total1} and \fig{fig:total2}. The main observation is that the effect of rotation increases with energy. One can qualitatively understand this dependence by noting that the classical trajectory of the fermion is a combination of two circular motions: one with angular velocity $\Omega$ due to the system rotation, and another one with angular velocity $\omega_B=qB/E$ due to the Lorentz force exerted by the magnetic field. The former is independent of the fermion energy $E$, whereas the latter decreases as $E^{-1}$. One can also notice that when the direction of rotation due to the magnetic field coincides with the direction of the system rotation of the fermion (e.g.\ $qB>0$ and $\Omega<0$), the result is enhancement of radiation. This happens because the rotating fermion experiences smaller effective $\omega_B$, hence smaller effective magnetic field \cite{Tuchin:2021lxl}. Conversely, when the two rotations are in the opposite direction (e.g.\ $qB>0$ and $\Omega>0$) we observe suppression of the radiation.  

At $\Omega=0$,  the quasi-classical formula (dashed line) approaches our exact result (dashed-dotted line) at high energy $E$. This is because the quasi-classical approximation neglects the discreteness of the fermion spectrum, which is a good approximation only in the ultra-relativistic case. It is remarkable that at high energy, the intensity of radiation by the rotating system deviates from that at $\Omega=0$ even for very small $\Omega$'s.  

It seems from our numerical results that the maximum of the ratio $W/W_\text{cl}$ for $\Omega<0$, or the inflection point for $\Omega>0$, depends on angular velocity roughly as $E_\text{max}\sim -\log_{10}|\Omega|-2$. If this trend persists at even lower $|\Omega|$'s, then the effect of rotation on the synchrotron radiation may be essential even in non-extreme astrophysical systems that rotate with typical angular velocities. This observation is a strong motivation to investigate the synchrotron radiation in a variety of magnetic fields and angular velocities and will be a subject of a further study. 

The effect of rotation on the synchrotron radiation that we have reported in this letter is mostly classical as it stems from the peculiar form of the metric in the rotating coordinates. We expect that the quantum effects induced by rotation become prominent when the angular velocity becomes comparable or larger than the inverse magnetic length.

In heavy-ion collisions, the direction of the magnetic field and the direction of rotation coincide. This implies that the synchrotron radiation by the negative charges must be significantly stronger than by the positive charges. As a result, we expect that rotation significantly enhances the contribution of the synchrotron radiation to the total photon spectrum, as compared to the non-rotating case \cite{Tuchin:2014pka}.  The formalism developed in this letter lays the foundation for the phenomenological applications that should be addressed in a dedicated work. 

In summary, we computed the effect of  rotation on the synchrotron radiation in the limit of relatively slow rotation. We argued that the effect of rotation is surprisingly strong which makes it amenable to experimental study.

%%%%%%%%%%%%%%%%%%%%%%%%%%%%%%%%
\bigskip
\acknowledgments
%I  am grateful to ... for many fruitful discussions of related problems. 
%We thank ... for helpful communications/correspondence.
This work  was supported in part by the U.S. Department of Energy under Grant No.\ DE-FG02-87ER40371.
%\clearpage
%%%%%%%%%%%%%%%%%%%%%%%%%%%%%%%%%%%%%

\end{document}